\documentclass{elsart}
\usepackage{graphics}
\usepackage{graphicx}
\usepackage{epstopdf}
\usepackage{epsfig}
\usepackage{amssymb}
\usepackage{makeidx}
\usepackage{amsfonts}
\usepackage{amstext}
\usepackage{amsmath}
\usepackage{amsbsy}
\usepackage{wasysym}
\usepackage[vflt]{floatflt}

\def\mip{\it{mip}\rm}
\def\WArP{\sf{WArP}\rm~}
\def\N2{N$_2$~}
\def\A{\kern+.6ex\lower.42ex\hbox{$\scriptstyle \iota$}\kern-1.20ex a}
\def\E{\kern+.5ex\lower.42ex\hbox{$\scriptstyle \iota$}\kern-1.10ex e}

\begin{document}

\begin{frontmatter}

\title{Oxygen contamination 
 in liquid Argon: combined effects on ionization electron charge and scintillation light}

\vspace*{-0.6cm}
{\large\sf WArP Collaboration}\\
{\small
\author[UAquila]{\small R.Acciarri}{,}
\author[LNGS,UAquila]{\small M.Antonello}{,}
\author[Padova]{\small B. Baibussinov}{,}
\author[UPadova]{\small M.Baldo-Ceolin}{,} 
\author[UPavia]{\small P.Benetti}{,}
\author[Princeton]{\small F.Calaprice}{,} 
\author[Pavia]{\small E.Calligarich}{,} 
\author[UPavia]{\small M.Cambiaghi}{,}
\author[UAquila]{\small N.Canci\corauthref{cor}}{,}
\author[UNapoli]{\small F.Carbonara}{,} 
\author[UAquila]{\small F.Cavanna\corauthref{cor}}{,}
\corauth[cor]{Corresponding authors}
\author[UPadova]{\small S. Centro}{,} 
\author[Napoli]{\small A.G.Cocco}{,}
\author[UAquila,LNGS]{\small F.Di Pompeo}{,}
\author[UNapoli]{\small G.Fiorillo}{,} 
\author[Princeton]{\small C.Galbiati}{,}
\author[Napoli]{\small V. Gallo}{,}
\author[LNGS,UAquila]{\small L.Grandi}{,}
\author[Padova]{\small G. Meng}{,}
\author[UAquila]{\small I.Modena}{,}
\author[Pavia]{\small C.Montanari}{,} 
\author[LNGS]{\small O.Palamara}{,} 
\author[LNGS]{\small L.Pandola}{,}
\author[Padova]{\small F. Pietropaolo}{,} 
\author[Pavia]{\small G.L.Raselli}{,} 
\author[Pavia]{\small M.Roncadelli}{,}
\author[Pavia]{\small M.Rossella}{,} 
\author[LNGS]{\small C.Rubbia}{,}
\author[LNGS]{\small E.Segreto}{,}
\author[Cracow,UAquila]{\small A.M.Szelc}{,}
\author[LNGS]{\small F.Tortorici}{,}
\author[Padova]{\small S. Ventura}{,}
\author[Pavia]{\small C.Vignoli}
}
\address[UAquila]{Universit\`a dell'Aquila e INFN, L'Aquila, Italy}
\address[LNGS] {INFN - Laboratori Nazionali del Gran Sasso, Assergi, Italy}
\address[Padova]{INFN - Sezione di Padova, Padova, Italy}
\address[UPadova]{Universit\'a di Padova e INFN, Padova, Italy}
\address[UPavia]{Universit\'a di Pavia e INFN, Pavia, Italy}
\address[Pavia]{INFN - Sezione di Pavia, Pavia, Italy}
\address[Princeton]{Princeton University - Princeton, New Jersey, USA}
\address[UNapoli]{INFN - Sezione di Napoli, Napoli, Italy}
\address[Napoli]{Universit\'a di Napoli e INFN, Napoli, Italy}
\address[Cracow]{IFJ PAN, Krakow, Poland}
\begin{abstract}
A dedicated  test of the effects of Oxygen contamination in liquid Argon has been performed at the 
INFN-Gran Sasso Laboratory (LNGS, Italy) within the \WArP R\&D program. Two detectors have been used: the \WArP 2.3 lt prototype and a small (0.7 lt) dedicated detector, coupled with 
a system for the injection of controlled amounts of gaseous Oxygen.\\
Purpose of the test with the 0.7 lt  detector is to detect the reduction of the long-lived component  lifetime of the Argon scintillation light emission at increasing O$_2$ concentration.  Data from the \WArP prototype are used for determining the 
behavior of both the ionization electron lifetime and the scintillation long-lived component  lifetime during the O$_2$-purification process activated in closed loop during the acquisition run. The electron lifetime measurements  allow to infer the O$_2$ content of the Argon and correlate it with the long-lived scintillation lifetime data.\\
The effect of Oxygen contamination on the scintillation light has been thus measured over a wide range of O$_2$ concentration, spanning from $\sim$10$^{-3}$ ppm up to $\sim$10 ppm. The rate constant of the light quenching process induced by Oxygen in LAr has been found to be $k'$(O$_2$)=0.54$\pm$0.03~$\mu$s$^{-1}$ppm$^{-1}$.

 \end{abstract}

 \end{frontmatter}

\tableofcontents

\vspace{-0.5cm}
\section{Introduction}
\label{sec:Introd}
Ionization in liquid Argon (LAr) is accompanied by scintillation light emission. 
Charged particles interacting in LAr create free electrons ($e^-$) and
excited Ar molecular states (Ar$^*_2$) which produce scintillation
radiation through de--excitation processes \cite{doke}.
Therefore, free electron separation and light emission are the two features
that characterize the use of LAr as active medium \cite{kubota}.
The two-phase (liquid/gas) technology developed by the \WArP Collaboration is based on the simultaneous collection of both signals \cite{warp1}.\\
The two processes are complementary and their relative weight depends on the 
strength of the electric field (EF) applied to the active LAr volume \cite{kubota}. The free electron yield  
(from ionization) rises with the field value while the photon yield (from scintillation) decreases.
For a minimum ionizing particle (\mip) at low fields (e.g. 0.5 kV/cm) the free electron yield is Y$_{ion}$ = $\sim$2.9$\times$ 10$^4$ $e^-$/MeV \cite{miyajima}. At the same EF 
the photon yield  is also high (Y$_{ph}$ =$\sim 2.4\times$ 10$^4$ $\gamma$/MeV) \cite{doke2}, for more densely ionizing particles is even higher. The light emission is characterized by a narrow--band intrinsic spectrum (around 127 nm) in the deep ultra-violet region (Vacuum-UV, VUV)  \cite{morikawa},\cite{grosjean} and  by a two-component exponential decay (short and long), characterized by $\tau_S$ in the 2 ns to 6 ns lifetime range and $\tau_T$ in the 1.1 $\mu$s to 1.6 $\mu$s  range respectively \cite{morikawa},\cite{hitachi}. 

Tiny fractions of impurities (O$_2$, N$_2$, H$_2$O and CO+CO$_2$) diluted at $\le$ 1 ppm level in LAr  are usually reported in commercially available Argon (best grade), due to the industrial process of air separation. These impurities may impair the detector performance by significantly reducing the  amount of both charge and light available from ionization events in LAr. 
This is due to various reactions involving different types of impurities, the most harmful of which being Oxygen. Since the pioneering R\&D work by the Icarus Collaboration \cite{buckley}, Argon purification systems (Oxygen reactants and molecular sieves) are known to be very effective in reducing the O$_2$ contamination and also H$_2$O and CO+CO$_2$ to a negligible level (down to $\le 0.1$ ppb). 
  
Detailed studies of the effects of Oxygen contamination on ionization electrons have been performed in the past. On the other hand, the effects on LAr scintillation light have been less investigated\footnote{A study of the effects of air traces (N$_2$+O$_2$) in Argon gas can be found in Ref.\cite{Amsler}.}. Purpose of the present work is to elucidate this last aspect, eventually by combining with available data on free electron trapping by O$_2$.

\vspace{-0.5cm}
\section{Residual Oxygen contamination in liquid Argon}
\label{sec:O2_contam}
\vspace{-0.5cm}
Two are the effects attributed to O$_2$ contamination in LAr: (1) the ionization electron attachment and (2) the scintillation light quenching. Characteristics of these processes are briefly summarized here below.

\vspace{-0.5cm}
\subsection{Ionization electron attachment}
\vspace{-0.5cm}
The main limitation for a full collection of the free electron charge in LAr is due to the residual concentration of electro-negative molecules. In fact, electron attachment processes to electro-negative impurities like O$_2$ are active:
\begin{equation}
e^-~+~O_2~\rightarrow O_2^-
\end{equation} 

Under the assumption that the free electron concentration [$e^-$] produced by ionizing events is smaller than the impurity concentration\footnote{The concentration units in use here are ppm, {\it parts per million (atomic,} often indicated as ppma) and ppb, {\it part per billion}. These  indicate the ratio between the number of interesting elements (in our case O$_2$ molecules) to ordinary elements (Ar atoms).} [O$_2$], the free electron concentration
decreases in time as:
\begin{equation}
\frac{d~[e^-]}{dt}~=~-~k_e~[O_2]~[e^-]~~~\Rightarrow~~~[e^-](t)~=~[e^-](0)~e^{-t/\tau_e}
\label{eq:e_attach}
\end{equation}
where the {\it electron lifetime} $\tau_e$ is defined as:
\begin{equation}
\frac{1}{\tau_e}~=~{k_e~[O_2]}
\label{eq;O2_conc}
\end{equation}
The value of the {\it rate constant} $k_e$ depends upon the drift field applied to the active LAr volume.
The reference value\footnote{Other  measurements of $k_e$ are reported in literature \cite{icarus} or under the value of correlated variables \cite{hofmann},\cite{biller}: these agree within $\sim\pm 25$\%. At null field a value of 1$\times 10^{11}$ lt~moles$^{-1}$~s$^{-1}$ is also reported in \cite{bakale}, corresponding to an electron lifetime of $\sim$ 3 ms at 0.1 ppb of Oxygen concentration.}
at EF=1 kV/cm is $k_e$= 5.5$\times 10^{10}$ lt~moles$^{-1}$~s$^{-1}$ \cite{bakale} (equivalent to 1.9~ppm$^{-1}$ $\mu$s$^{-1}$).
\vspace{-0.5cm}
\subsection{Scintillation light quenching}
\vspace{-0.5cm}
 Residual O$_2$ contamination leads also to a substantial reduction of the 
scintillation light intensity.
This may be attributed to a quenching process in two-body collision of O$_2$ impurities with Ar$^*_2$ excimer states:
\begin{equation}
Ar^*_2~+~O_2~\rightarrow~ 2 Ar~+~O_2
\label{eq:O2_quench}
\end{equation}
This non-radiative collisional reaction is in competition with the de-excitation process leading to VUV light emission. As a result, a  sensitive quenching of the scintillation light yield is expected, mainly in the slow-component amplitude.
In fact, the quenching process leads to a decrease of the excimer concentration [Ar$^*_2$], while the 
contaminant concentration  [O$_2$] stays constant in time. To a first approximation, also for this case a 
first order rate law can be assumed, characterized by the rate constant $k'$:
\begin{equation}
\frac{d~[Ar^*_2]}{dt}~=~-~k'~[O_2]~[Ar^*_2] ~~~\Rightarrow~~~[Ar^*_2](t)~=~[Ar^*_2](0)~e^{-t~k'~[O_2]}
\label{eq:1st_ord_law}
\end{equation}
The slow component lifetime of the scintillation light emission results to be effectively decreased, depending on the O$_2$ concentration:
 \begin{equation}
 \frac{1}{\tau'_{T}}~=~\frac{1}{\tau_{T}}~+~k'~[O_2]
  \label{eq:tau_Qj}
 \end{equation}
To our knowledge, no data are reported in literature regarding the actual value of the Oxygen rate constant {\it k'} and its measurement is one of the goals of the present study.\\

\vspace{-1.5cm}
\section{\WArP 2.3 lt prototype test}
\vspace{-0.5cm}
The \WArP prototype is operated underground at the Gran Sasso INFN Laboratory (LNGS), in the tunnel connecting HallA to HallB. Data reported here refer to the Aug. '07 run.
The detector is a two-phase drift chamber, with a lower liquid
Argon volume and an upper region with Argon in the gaseous phase, both
viewed by the same set of photo-multipliers (PMTs) \cite{warp2}. Free electrons 
generated by ionization in the liquid are drifted by means of an electric field to the
liquid-gas interface, where they are extracted through the boundary and
detected by the proportional scintillation light generated by the electrons
accelerated in a high electric field.\\
The drift volume, 7.5 cm long, is delimited by a 20 cm diameter stainless steel
cathode and by a system of field-shaping electrodes that generate very
uniform electric drift fields in the 1.87 lt sensitive
volume ($\simeq$ 2.3 lt total volume). A grid (g1) placed just below the
liquid level closes the uniform drift field region, while two additional grids
(g2 and g3 from bottom to top) are placed in the gas phase.
In normal data taking conditions, the chamber is operated with a drift field of
EF~=~1 kV/cm.\\
Four 12-stage 3" photo-multipliers (PMTÕs), manufactured in order to operate at LAr
temperature and placed at about 4 cm above the last grid, detect both the
primary scintillation in the liquid (the S1 signal) and the proportional scintillation light in gas phase (S2 signal). 
Sensitivity to VUV photons emitted by the scintillating Argon is achieved by coating the
photo-multiplier window with an appropriate compound, i.e. Tetra-PhenylButadiene (TPB), 
which acts as a fluorescent wavelength shifter of the VUV
scintillation light to the photo-multiplier sensitive spectrum. 
In order to improve the light collection efficiency from the drift volume, a
high performance diffusive reflector layer with TPB deposit 
surrounds the inner drift volume
and the gas volume between the top grid and the PMTÕs. \\
The system is contained in a stainless steel, vacuum-tight cylindrical vessel. 
The whole container is cooled down to
about 86.5 K by an external liquid Argon bath. This set-up ensures in the
inner container a constant absolute pressure few mbar above the external
atmospheric pressure (P$\simeq 900$ mbar).\\
Gaseous Argon (GAr) for filling is the best grade 6.0 (99.99990\%) commercial Argon, with impurity concentration below 1 ppm
(of which $\le$ 0.2 ppm of O$_2$ and $\le$ 0.5 ppm of Nitrogen, N$_2$, according to specifications).
At  filling, GAr is flushed through a standard Oxysorb/Hydrosorb cartridge for initial removal of O$_2$ and H$_2$O.
 An Argon recirculation system is implemented working in closed loop and providing a further continuous
re-purification of the Argon contained in the chamber. \\
The anode signal from  of each PMT is integrated (shaping time 120 $\mu$s) and sent to
a 10 bit flash ADC with 100 MHz sampling frequency. At each trigger the memory buffer is then
recorded to disk.\\

\vspace{-1.cm}
\subsection{Electron lifetime measurement}
\label{sec:e_lifetime}
\vspace{-.5cm}
The amount of free $e$-charge released by ionization events in the liquid phase can be determined by the pulse-height of the secondary signal (S2). This is proportional to the number of free electrons
extracted in the gas phase. According  to the reaction of Eq.\ref{eq:e_attach}, electron attachment to electro-negative O$_2$ (or O$_2$-equivalent) molecules diluted in LAr may occur during the drift up to the liquid-to-gas separation surface. Therefore, the amount of charge extracted results to be exponentially lessened at increasing drift distance of the ionization event from the surface. \\
\begin{figure}[h]
\begin{center}
\vspace{-0.8cm}
\includegraphics[width=9.2cm,angle=0]{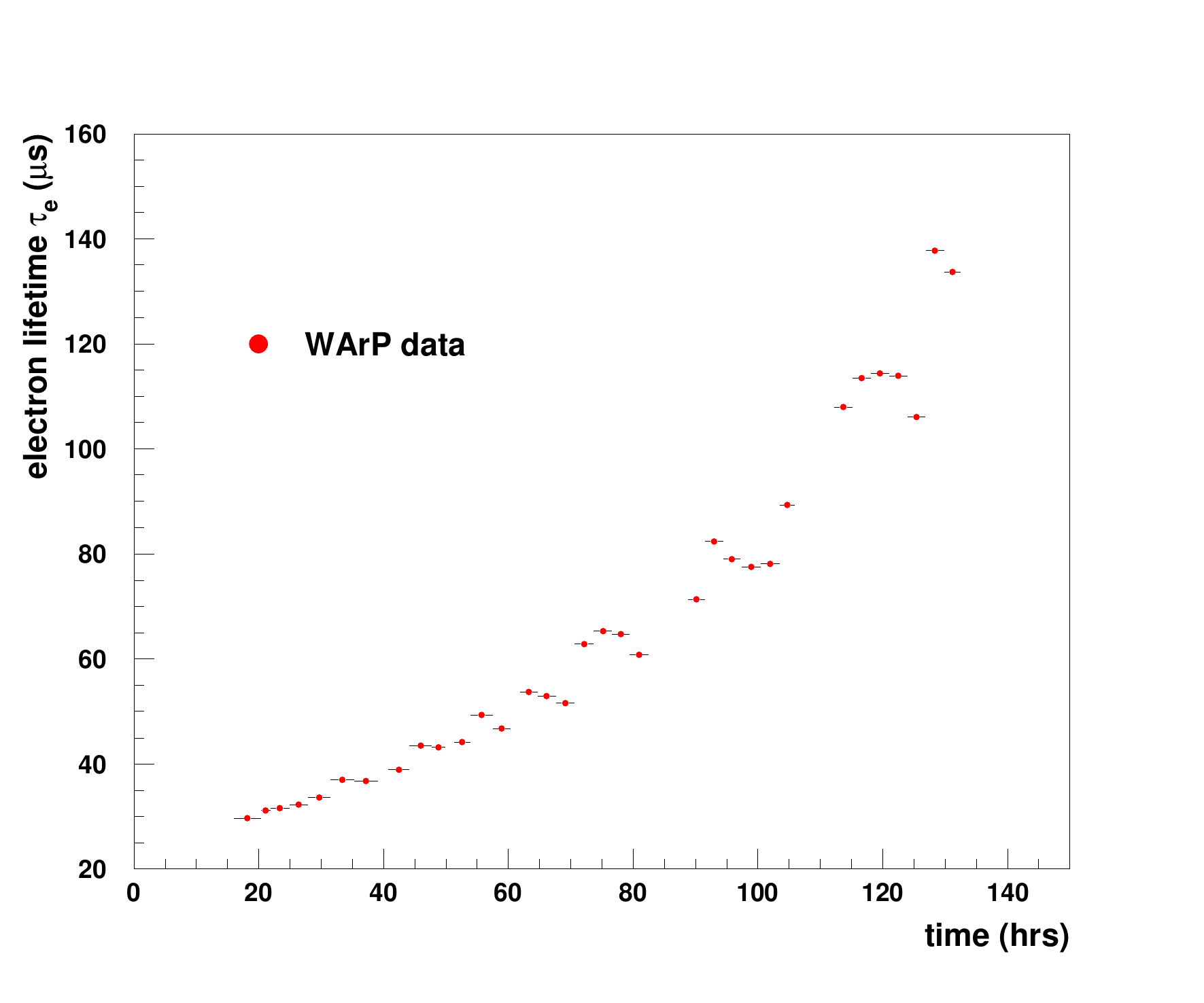}
\vspace{-0.5cm}
\includegraphics[width=9.2cm,angle=0]{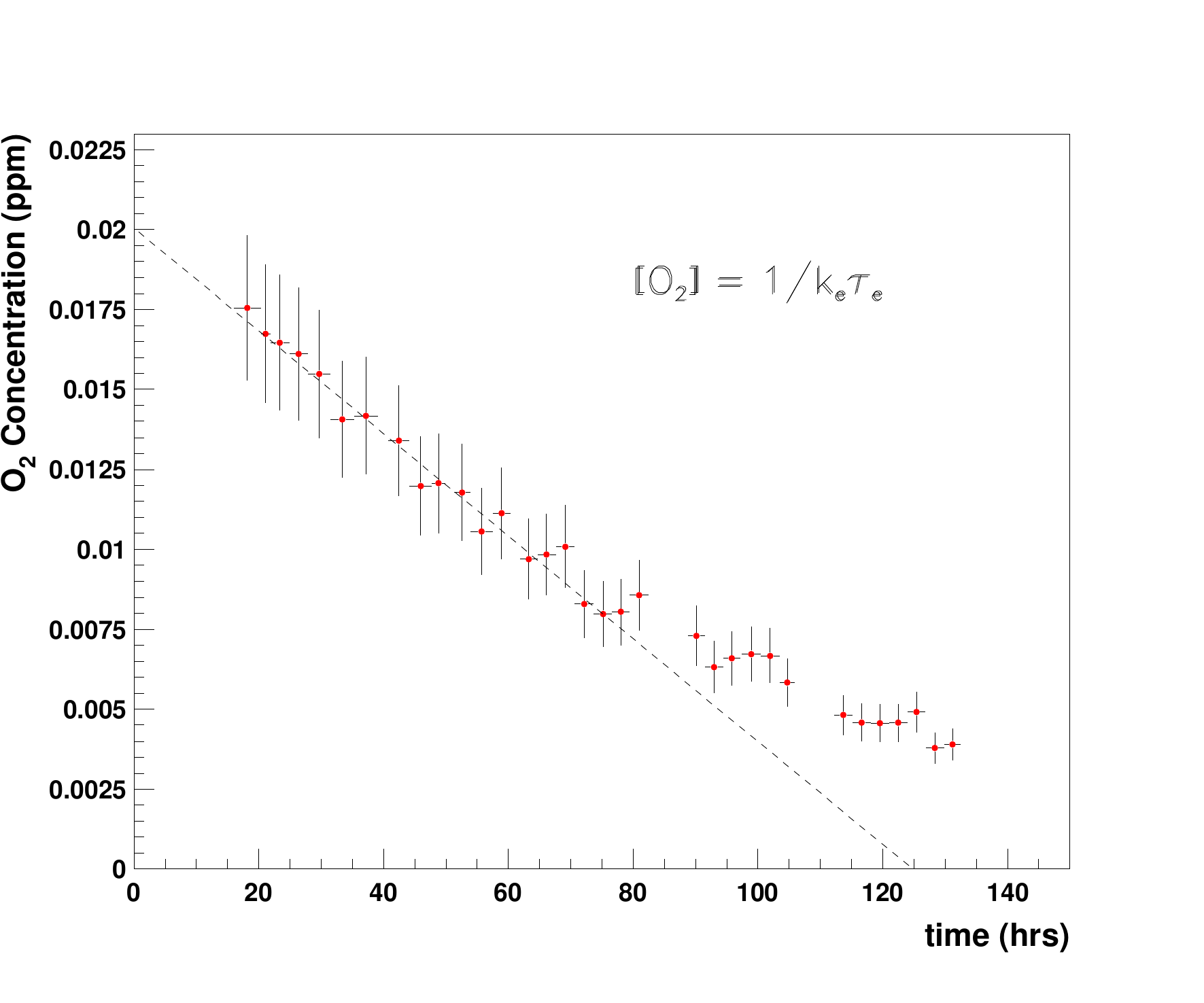}
 \caption{\textsf{\textit{Time evolution of  $\tau_e$ [Top] and O$_2$ concentration [Bottom] in the WArP 2.3 lt chamber during Argon purification process.}}}
\label{fig:cont}
\end{center}
\end{figure}
For each recorded ionization event the drift time can be determined as the time interval between S2 and S1 signals.  The distribution of the S2/S1 pulse-height ratio vs drift time for the events\footnote{These are background events mainly from beta emission of $^{39}$Ar in natural Argon.} recorded in a defined time slot shows an exponential decay trend, with characteristic time constant  (the {\it electron lifetime} $\tau_e$) to be determined by fitting. The corresponding O$_2$ concentration can be inferred from Eq.\ref{eq;O2_conc}, through the $k_e$ value at the drift field of operation (assumed as known \cite{bakale}, 1.9~ppm$^{-1}$ $\mu$s$^{-1}$).\\
Measurements performed in successive time slots (of about 3 hrs) after filling, during Argon recirculation through 
the purification system, indicate that the electron lifetime progressively increases (i.e. the Oxygen concentration reduces) up to the limit of sensitivity of the method employed ($\ge$ 200 $\mu$s). 
The results are shown in Fig.\ref{fig:cont}, [Top] the measured $\tau_e$ and [Bottom] the corresponding [O$_2$] behavior as a function of run time after filling.
The decrease of the O$_2$ concentration initially follows a linear behavior in time; this allows to extrapolate the initial O$_2$ contamination of the Argon in the chamber. This is found to be around 20 ppb. It comes from the reduction of the original O$_2$ content in the 6.0 grade GAr ($\le$ 0.2 ppm), due to the purification procedure at filling time, partially balanced by contributions from unavoidable material outgassing inside the detector and from the residual leak rate of the filling system. \\

\subsection{Long-lived scintillation lifetime measurement at low O$_2$ concentration}
\label{sec:tau_T1}
\begin{figure}[h]
\begin{center}
\includegraphics*[width=11.5cm,angle=0]{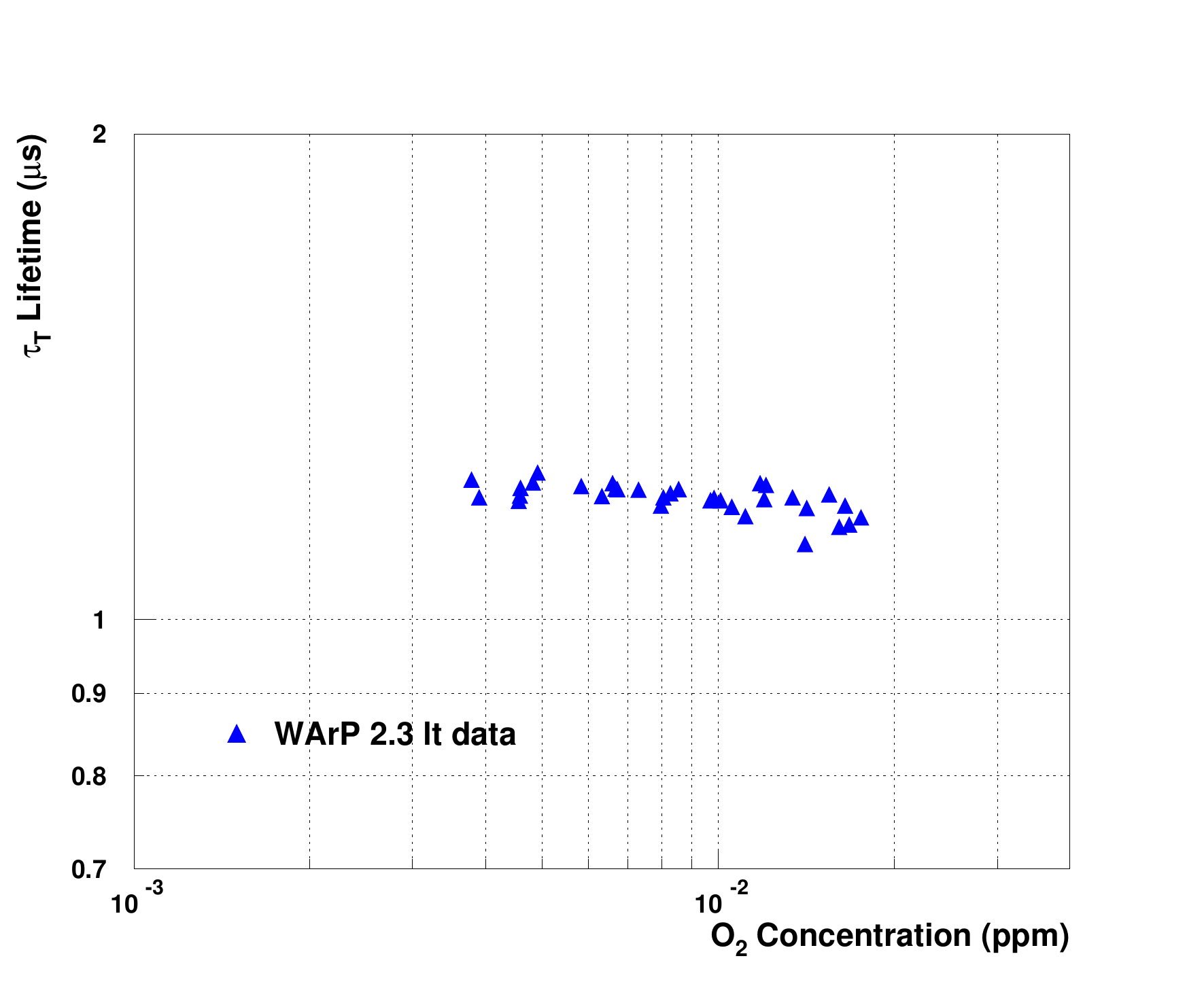}
 \caption{\textsf{\textit{Long-lived scintillation lifetime from the fit of the average waveforms (WArP 2.3 lt data) vs Oxygen concentration (inferred from the electron lifetime analysis).}}}
\label{fig:taue_tauT}
\end{center}
\end{figure}

The PMT signal with the \WArP 2.3 lt detector is integrated and digitized with 10 ns sampling time. Pulse shape analysis of the recorded primary signals (S1) allows the reconstruction of the 
main features of the scintillation light emission following ionization events in LAr.  In particular, taking into account the shaping time of the charge amplifier in use, the long-lived component decay time ($\tau_T$)  can be determined by a fitting procedure of the signal shape. 
The fit is applied to the waveform obtained as the average of the signals recorded in each time slot defined in Sec.\ref{sec:e_lifetime}. In Fig.\ref{fig:wfms} [Left] an averaged waveform (from data in a given time slot) is shown with the fitting function superimposed.\\
The [O$_2$] mean level of the corresponding time slot has been inferred from the electron lifetime analysis (Sec.\ref{sec:e_lifetime}). The correlation of the measured long-lived decay time ($\tau_T$) with the actual Oxygen concentration is reported in Fig.\ref{fig:taue_tauT} for the full set of data.\\
The correlation is rather weak and the almost constant $\tau_T$ behavior in the range below $\simeq$ 20 ppb indicates that at this (low) contamination level scintillation emission is unaffected (or only marginally affected) by  the quenching process of Eq.\ref{eq:O2_quench}. Dedicated measurements at higher O$_2$ concentrations are needed to find appreciable effects (see Sec.\ref{sec:tau_T2}).\\

\vspace{-1.2cm}
\section{Oxygen contamination test with a 0.7 lt detector}
\label{sec:LAr_Det}
\vspace{-0.5cm}
A  detector  dedicated to the study of Oxygen contamination effects on LAr scintillation light has been assembled and operated at the LNGS external facility (``Hall di Montaggio'', Jan.-Apr. '07). It consists of a cylindrical LAr cell in PTFE (h=12 cm, $\varnothing$=8.5 cm internal dimensions, 
wall thickness 0.5 cm) containing about 0.7 lt of LAr (about 1 kg of active mass), viewed by a single 2" 
PMT mounted on the open top end of the  cell, of the same type of those in use with the \WArP prototype (including the wave-length shifting coating of TPB). 
A reflector layer also covered with TPB surrounds the internal walls (side and bottom) of the PTFE cell.\\
The detector is housed in a stainless steel cylindrical chamber, closed at both ends by vacuum tight  flanges. The internal volume of the chamber is about 6.5 lt and it contains, after filling, 
a total amount of 3.0 lt of LAr (including the LAr cell active volume at its bottom).  The chamber is immersed in a LAr bath of a stainless steel open dewar, to liquify and keep at stable temperature the LAr internal volume
(T$\simeq 86.7$ K, P$\simeq 910$ mbar).\\
A transfer line for the Ar filling and for the injection of controlled amounts of O$_2$ (from a minimum of about 50 ppb per injection) has been connected to the LAr chamber. Gaseous Argon (GAr) for filling is the best grade 6.0 commercial Argon, as for the \WArP 2.3 lt run. 
The GAr is flushed through a Hydrosorb/Oxysorb cartridge for ``partial" removal of O$_2$ and H$_2$O positioned along the GAr filling line. No additional recirculation/O$_2$-repurification system is implemented with this detector.\\
The PMT anode signal is directly read--out by a fast {\it Waveform Recorder} with sampling time of 1 ns  over a full record length of 10 $\mu$s.\\
After the first run at 0 ppm (i.e. no additional O$_2$ 
injection to the initial purified 6.0 Argon), an experimental test was performed by adding progressively controlled amounts of O$_2$ (total content 60 ppb, 300 ppb, 600 ppb, 2 ppm, 5 ppm, 10 ppm), waiting for few hours and exposing the LAr cell to $\gamma$-sources after each contamination.\\
More information about the detector and details of the data treatment can be found in \cite{N2_test} where results from another test of LAr contamination with Nitrogen are reported.

\vspace{-0.4cm}
\subsection{\vspace{-0.6cm}Long-lived scintillation lifetime measurement at high O$_2$ concentration}
\label{sec:tau_T2}
Waveform recording of the direct PMT signal (no charge amplifier stage) allows for a detailed study of the LAr scintillation light emission, in particular of the individual exponential components (relative amplitude and decay time).\\
To this purpose the average waveforms, from each run taken in presence of incremental O$_2$ contamination with exposure to a $^{60}$Co $\gamma$-source, have been off-line processed to obtain the corresponding {\it scintillation signal shapes} according to the procedure defined in \cite{N2_test}.\\
\begin{figure}[b]
\begin{center}
\vspace{-0.8cm}
\includegraphics*[width=10.5cm]{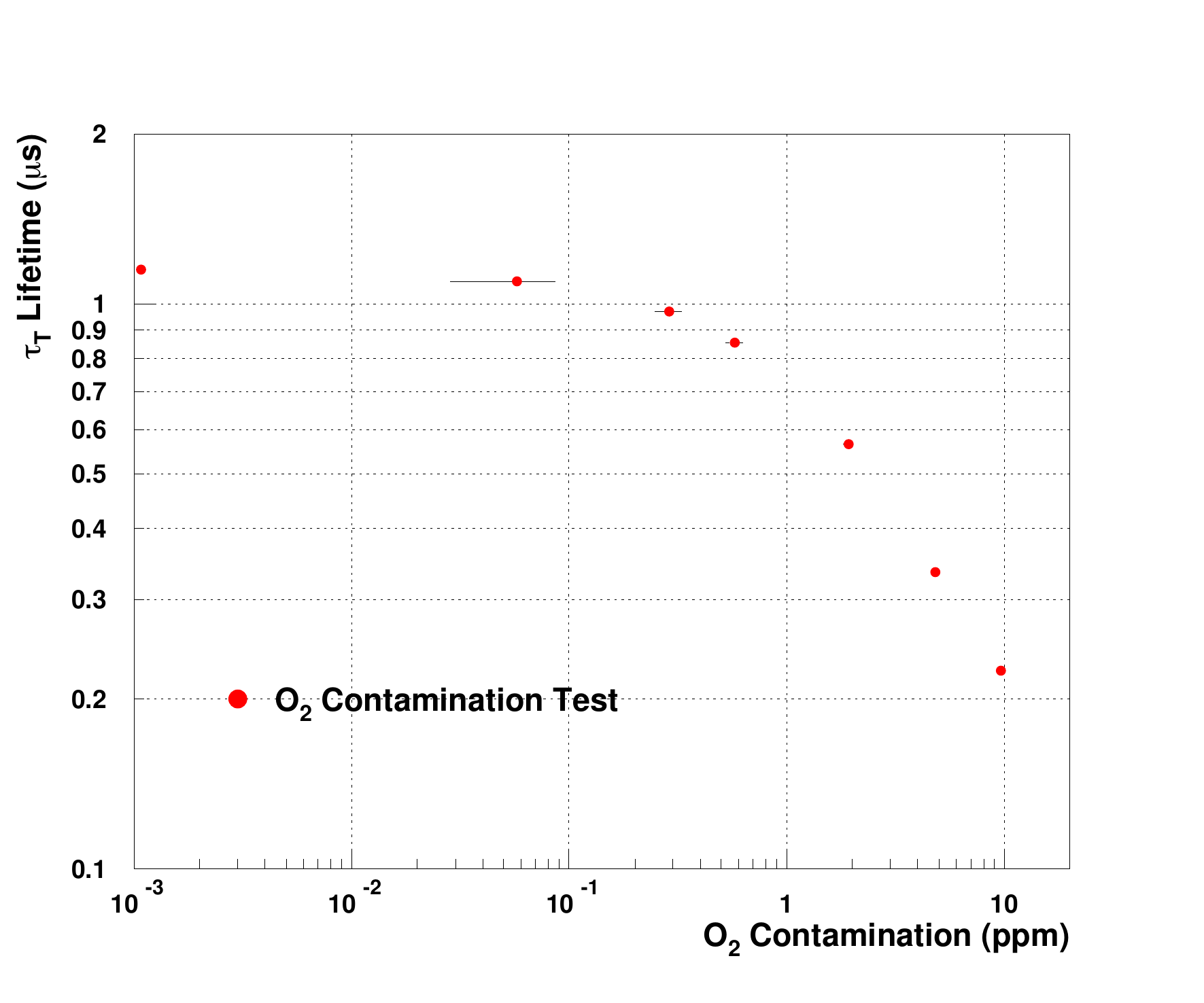}
\vspace{-0.6cm}
 \caption{\textsf{\textit{Long-lived scintillation lifetime at the various [O$_2$] contaminations (from 0.7 lt detector data).}}}
\label{fig:tauT_O2}
\end{center}
\end{figure}
By a fitting procedure of the signal shapes, the long-lived scintillation lifetime $\tau_T$ value has been determined. In Fig.\ref{fig:wfms} [Right] three signal shapes (runs at 0 ppm, 0.6 ppm and 2 ppm of O$_2$ contamination) are shown with the fitting functions superimposed.\\
The $\tau_T$ values from the fit of each run are reported in Fig.\ref{fig:tauT_O2}.\\
The data in Fig.\ref{fig:tauT_O2} show a clear decreasing behavior of $\tau_T$ as the O$_2$ contamination exceeds the $\sim$ 100 ppb level, due to the quenching process of Eq.\ref{eq:O2_quench}.\\
The lifetime of the fast component ($\tau_S$), also determined by the fitting procedure of the signal shape, is instead almost unaffected even at the largest contaminations. 

\vspace{-0.3 cm}
\subsection{\vspace{-0.3 cm}Signal Amplitude Analysis}
\begin{figure}[h]
\begin{center}
\vspace{-0.5cm}
\includegraphics*[width=10cm]{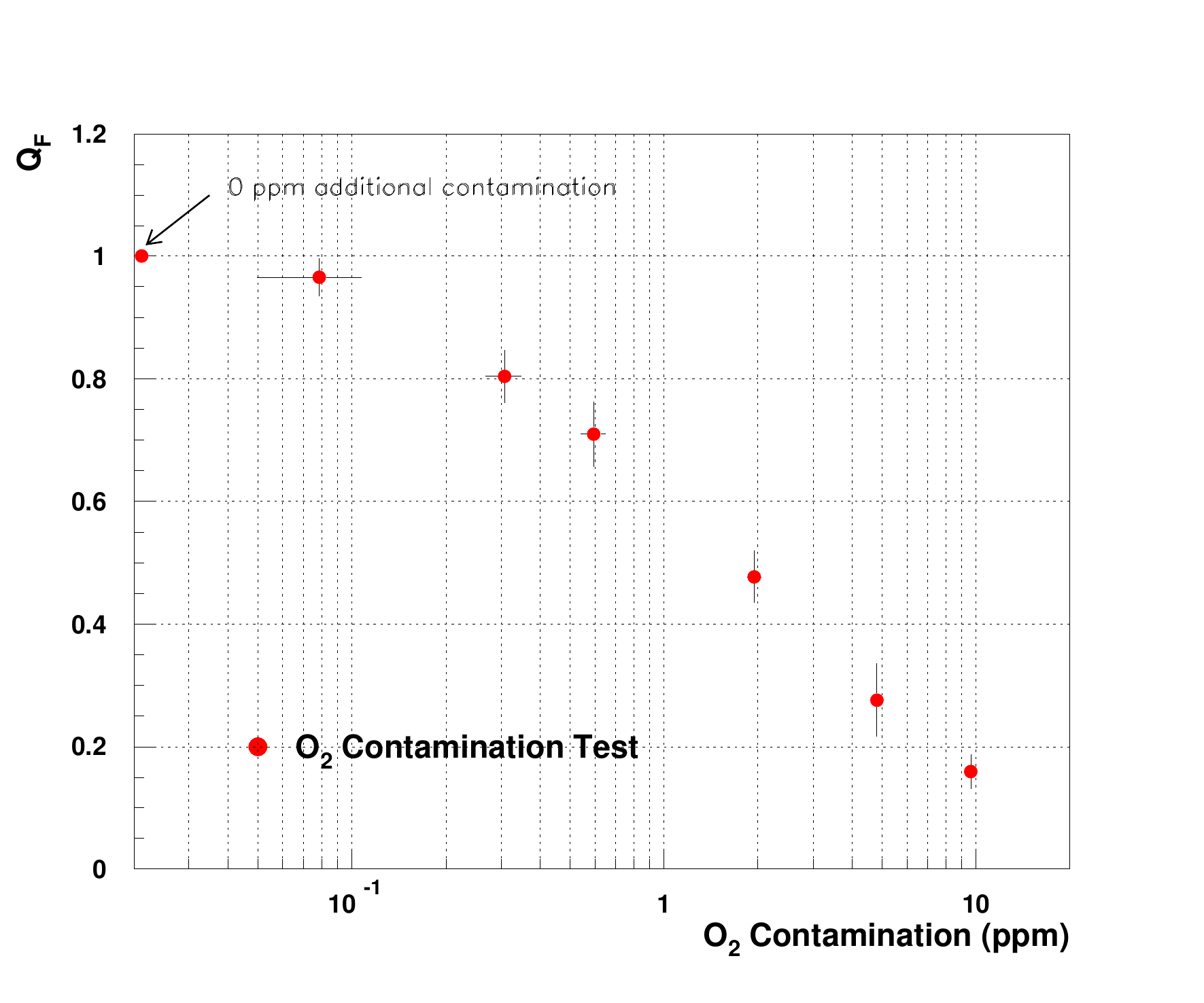}
\vspace{-0.5cm}
 \caption{\textsf{\textit{Quenching factor $Q_F$  behavior at incremental O$_2$ contaminations from pulse amplitude spectra analysis (~$^{60}$Co runs with the 0.7 lt test detector).}}}
\label{fig:QF_tot}
\end{center}
\end{figure}

The signal amplitude is obtained by single waveform integration. This amplitude, expressed in photo-electron units, is proportional to the energy deposited by electrons from $^{60}$Co source $\gamma$-conversion. 
Pulse amplitude spectra (mainly from Compton scattering) have been obtained for each run at different 
[O$_2$] value. These spectra result to be progressively down-scaled at increasing O$_2$ concentrations as due to the quenching process (\ref{eq:O2_quench}).\\
A dedicated fitting procedure has been developed to determine the quenching {\it scale factor}\footnote{The quenching factor $Q_F$ can be equivalently defined as the ratio between the total intensity of scintillation light emitted for a given O$_2$ contamination 
with respect to the case of pure liquid Argon.} Q$_F$ relative to each O$_2$ contamination: the individual entry values of the uncontaminated (0 ppm) spectrum are down-scaled by Q$_F$, free parameter, to best fit the experimental [O$_2$] contaminated spectra. The Q$_F$ values thus obtained are reported in Fig.\ref{fig:QF_tot} as a function of the O$_2$ contamination.  

A sensitive reduction of the scintillation light yield is found in this high concentration range ([O$_2$]$\ge$ 100 ppb), e.g. a loss of $\sim$ 40\% of the light available from scintillation is accounted to the presence of 1 ppm of O$_2$ in Argon. \\
This confirms that the use of appropriate O$_2$ purification systems is needed also for LAr detectors based on the collection of the scintillation light. 

\section{Combined analysis of the long-lived component lifetime}
\label{sec:comb_anal}
\begin{figure}[ht]
\begin{center}
\vspace{-0.5cm}
\includegraphics*[width=6.7cm]{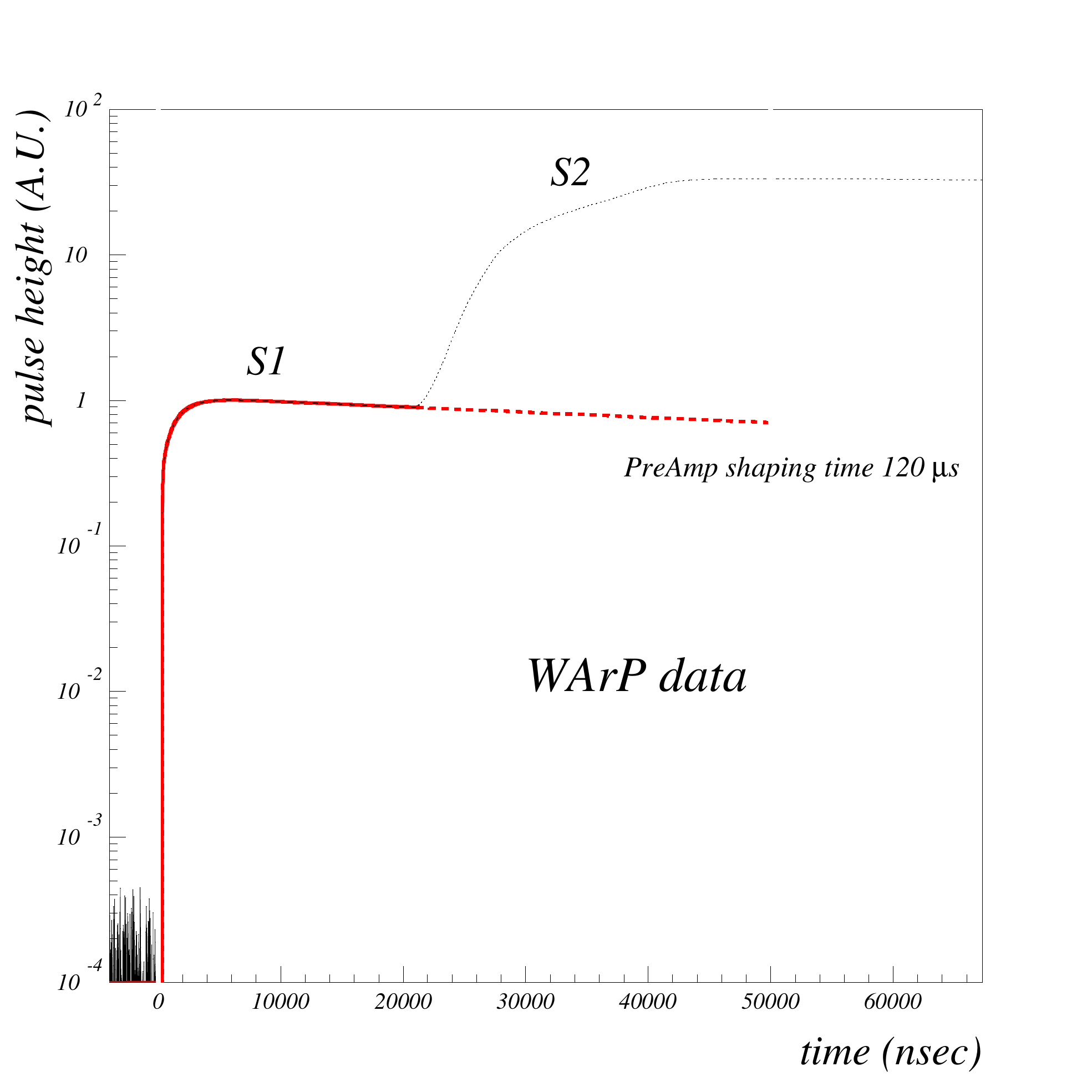}
\includegraphics*[width=6.7cm]{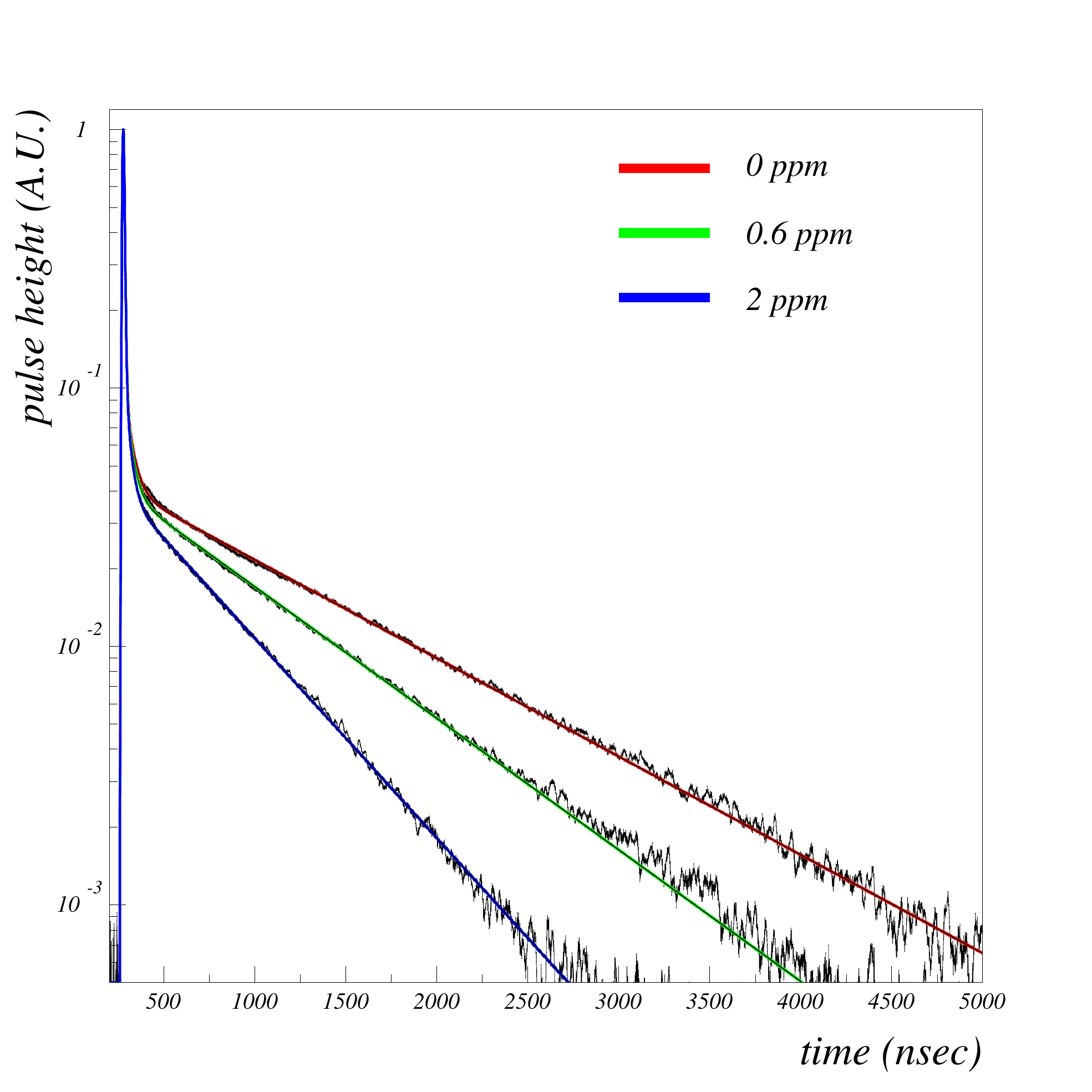}
\vspace{-0.3cm}
 \caption{\textsf{\textit{Examples of (averaged) signal waveforms collected with the WArP 2.3 lt prototype [Left] and with the 0.7 lt test detector [Right].}}}
\label{fig:wfms}
\end{center}
\end{figure}

\begin{figure}[hbt]
\begin{center}
\includegraphics*[width=13cm]{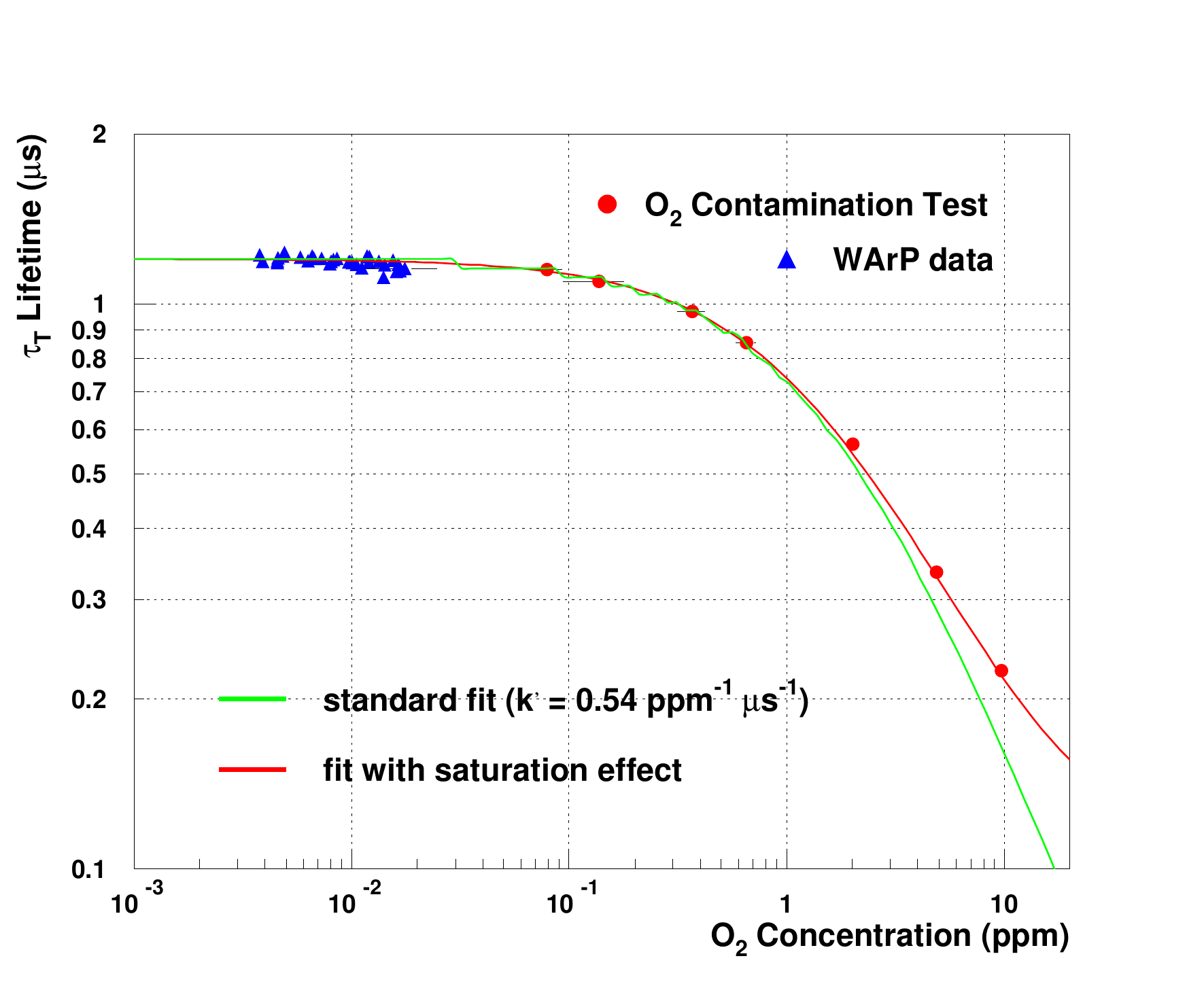}
 \caption{\textsf{\textit{Long lived scintillation lifetime $\tau_T$ vs O$_2$ concentration from the combined analysis of the WArP 2.3 lt prototype and of the 0.7 lt detector data.}}}
\label{fig:comb_fit}
\end{center}
\end{figure}

In the present study the long-lived lifetime $\tau_T$ characterizing the scintillation signal shape in LAr has been measured over a wide range of O$_2$ concentrations, spanning from $\sim$10$^{-3}$ ppm  (\WArP 2.3 lt data, e.g. Fig.\ref{fig:wfms} [Left]) up to $\sim$10 ppm (``contamination test" with the dedicated experimental set-up, 0.7 lt LAr detector, e.g. Fig.\ref{fig:wfms} [Right]).
All  the $\tau_T$ measurements are reported together in Fig.\ref{fig:comb_fit}. An overall $\chi^2$-minimization, using Eq.\ref{eq:tau_Qj} as fitting function (green curve in Fig.\ref{fig:comb_fit}), allows to extract the following results:
\begin{itemize}
\item the  rate constant of the quenching process (Eq.\ref{eq:O2_quench}):\\ $k'$(O$_2$)~=~0.54~$\pm$ 0.03 $\mu$s$^{-1}$~ppm$^{-1}$;
\item the slow-component lifetime at 0 ppb of Oxygen equivalent concentration:  $\tau_T~=~1.21~\pm ~0.01~\mu$s;
\item the initial O$_2$ concentration in the 0.7 lt detector run:\\ 
$[$O$_2$]$_{in}$~=~65~$\pm$~15 ppb\\
(this value is obtained by using an additive free parameter in the fitting function: [O$_2$]$\rightarrow$[O$_2$]+[O$_2$]$_{in}$).
\end{itemize}
The errors associated to the results are from the overall fit (statistical), taking into account the statistical errors from the fit of the signal shapes at different O$_2$ contaminations and also the error associated to the injected amounts of Oxygen in the contamination procedure.

The fitting behavior (green curve in Fig.\ref{fig:comb_fit}) compared to data shows an increasing discrepancy at higher O$_2$ concentrations, as possibly due to incomplete absorption of impurities into the liquid. This trend might be explained by a {\it saturation} effect of the solute (Oxygen) in the LAr solvent. Similar indications have been reported by other groups \cite{biller}. The relatively short time scale (hours to day) of the measurements performed prevents to draw any definite conclusion on long term effects.\\
A different parameterization of the fitting function for the slow component lifetime has been used, depending on the O$_2$ concentration under saturation hypothesis:
 \begin{equation}
 \frac{1}{\tau'_{T}}~=~\frac{1}{\tau_{T}}~+~k'~\beta~(1~-~e^{-\frac{[O_2]}{\beta}})
  \label{eq:tau_T_sat}
 \end{equation}
where the $\beta$ parameter represents the concentration scale where saturation becomes effective ($\beta=13$ ppm from fit) and 
at first order it reduces to the original fitting function (Eq.\ref{eq:tau_Qj}).\\
The red curve in Fig.\ref{fig:comb_fit} shows the result of the fit with the modified parameterization. This model considerably improves the scaling with O$_2$ concentration and decreases the overall $\chi^2$, without changing at all the results reported above.\\
 The goodness of the overall fit (saturation model) is satisfactory (C.L.$\simeq$90\%).

Residual concentration of non-electronegative contaminants in LAr cannot be excluded, i.e. [N$_2$] up to a maximum level of 0.5 ppm (according to the specifications of the 6.0 GAr used for filling in both experimental set-up's here employed). Scintillation light quenching process in LAr due to Nitrogen contamination is in fact known to be rather effective (e.g. with rate constant $k$(N$_2$)=0.11 $\mu$s$^{-1}$ppm$^{-1}$ \cite{himi}). This allows to compute an upper limit for the long-lived component lifetime assuming the N$_2$ concentration at its maximum level. The range of variability is thus defined as $1.21~\mu$s~$\le~\tau_T~\le~1.28~ \mu$s.\\
A test dedicated to the Nitrogen effect on LAr scintillation light has been successively performed with the 0.7 lt detector. The rate constant $k$(N$_2$) and the actual Nitrogen content in the 6.0 Ar have been measured as reported in \cite{N2_test}.

\vspace{-0.3 cm}
\section{\vspace{-0.3 cm}Conclusion}
Free electron charge separation and light emission are the two features
that characterize the use of LAr as active medium for ionization events.\\
Residual content of Oxygen in LAr  may impair the detector performance by significantly reducing the  amount of both charge and light available. \\
Detailed studies of the consequences of Oxygen contamination on ionization electrons have been reported in literature, showing dramatic effect even in extremely low concentrations as a result of its high electron affinity. This led to the development of appropriate purification systems, key element in the LAr-TPC technology.\\
The effects on scintillation light have been however less investigated.
Residual O$_2$ contamination leads to a substantial reduction also of the 
scintillation light intensity. This is due to a quenching (i.e. non-radiative) process in two-body collision of O$_2$ impurities with Ar$^*_2$ excimer states, in competition with the de-excitation process leading to VUV light emission. \\
Two detectors have been used for a detailed test, here reported, of the effects of O$_2$ concentration in LAr: the \WArP 2.3 lt prototype and a small (0.7 lt) dedicated detector, coupled with 
a system for the injection of controlled amounts of gaseous Oxygen.\\
With the 0.7 lt  detector the reduction at increasing O$_2$ concentration of the long-lived component  lifetime of the scintillation light emission has been studied, while data from the \WArP prototype were used for determining the behavior of both the ionization electron lifetime and the scintillation slow-component  lifetime during the O$_2$-purification process activated in closed loop during the acquisition run. The electron lifetime measurements  allow to infer the O$_2$ content of the Argon and correlate it with the long-lived scintillation lifetime data.\\
The effect of Oxygen contamination on the scintillation light has been thus measured over a wide range of O$_2$ concentration, spanning from $\sim$10$^{-3}$ ppm up to $\sim$10 ppm. The rate constant of the light quenching process induced by Oxygen in LAr has been found to be $k'$(O$_2$)~=~0.54~$\pm$~0.03~$\mu$s$^{-1}$ppm$^{-1}$. This value is large, e.g. compared with the corresponding value of the Nitrogen rate constant, indicating that the use of appropriate O$_2$ purification systems is needed also for detectors based on the collection of the LAr scintillation light. 

\section{Acknowledgments}
\label{sec:Acknow}
We would like to warmly thank the LNGS technical staff
for the valuable support during the various phases of the experimental test.\\
This work has been supported by {\it INFN} Istituto Nazionale di Fisica Nucleare, Italy), 
by {\it MIUR} (Ministero dell'Istruzione, dell'Universit\'a e della Ricerca, Italy) - Research Program Prot. 2005023073-003 (2005), 
by the {\it ILIAS} Integrating 
Activity (Contract RII3-CT-2004-506222) as part of the {\it EU FP6} programme in
Astroparticle Physics, by a grant of the President of the 
Polish Academy of Sciences and by {\it MNiSW} grant 1P03B04130.

\clearpage


\end{document}